\documentclass[12pt]{article}
\usepackage{amssymb,amsmath}
\usepackage[noblocks]{authblk}
\usepackage[top=0.75in, bottom=0.75in, left=0.75in, right=0.75in, dvips]{geometry}
\usepackage{caption}
\begin{document}
\textwidth 10.0in 
\textheight 9.0in 
\topmargin -0.60in
\title{Ghosts from Ghosts in the BRST Formalism}
\author[1,2]{D.G.C. McKeon}
\affil[1] {Department of Applied Mathematics, The
University of Western Ontario, London, ON N6A 5B7, Canada} 
\affil[2] {Department of Mathematics and
Computer Science, Algoma University, Sault St.Marie, ON P6A
2G4, Canada}
\date{}
\maketitle

\maketitle
\noindent
email: dgmckeo2@uwo.ca\\
PACS No.: 11.10Ef\\
KEY WORDS: BRST, ghosts

\begin{abstract}
We show that the Hamiltonian $H_Q$ introduced in the course of BRST analysis of a gauge theory may in fact be associated with an action that itself is gauge invariant.  This action can then be treated using the BRST formalism.  We illustrate this by considering the spinning particle and the first order Einstein-Hilbert action in $1 + 1$ dimensions.
\end{abstract}

\section{Introduction}

The treatment of systems whose action involves non-physical degrees of freedom through the Hamiltonian-BRST procedure [1-3] is quite useful. (For reviews, see refs. [4-7].)  In this approach, the contribution of non-physical degrees of freedom to physical processes is cancelled by systematically introducing additional non-physical degrees of freedom with opposite statistics.

The first step in this procedure is to introduce a canonical pair of ``ghosts'' ($\theta_i, \pi_i$), one for each first class constraint $\phi_i$ arising in the theory, and having different Grassmann character from $\phi_i$.  Next, a BRST operator $Q$ is introduced such that
\begin{equation}
Q = \theta_i\phi_i + Q_E 
\end{equation}
where $Q_E = Q_E (q_i, p_i, \theta_i, \pi_i)$ is the ``extra'' contribution to $Q$ that ensures that
\begin{equation}
\left\lbrace Q, Q \right\rbrace^\ast = 0
\end{equation}
where $\left\lbrace , \right\rbrace^\ast$ is a Dirac Bracket (DB) used to eliminate any second class constraints [5, 7, 8] that are present. Once $Q$ has been found, a ``BRST Hamiltonian'' $H_Q$ constructed so that 
\begin{equation}
\left\lbrace Q, H_Q \right\rbrace^\ast = 0
\end{equation}
with the condition that if the ghost fields were to all vanish, then $H_Q$ reduces to $H_C$, the canonical Hamiltonian.

In this paper, we wish to note that the BRST action
\begin{equation}
S_Q = \int d\tau \left(p_i \dot{q}_i + i\pi_i  \dot{\theta}_i - H_Q \right)
\end{equation}
may itself possess gauge symmetries, much like the classical action
\begin{equation}
S_C = \int d\tau \left(p_i \dot{q}_i - H_C \right).
\end{equation}
These symmetries can be found by either examining the equations of motion that follow from the action [9] or by examining the action itself [10].  In both approaches, a symmetry generator $G$ gives rise to a change in any dynamical variable $A$
\begin{equation}
\delta A = \left\lbrace A, G \right\rbrace^\ast
\end{equation}
that leaves the action invariant.

We will consider two models in order to demonstrate how $S_Q$ can possess gauge symmetries.  First of all we shall examine the spinning particle [11] which possesses both a local ``Bosonic'' (non-Grassmann) and ``Fermionic'' (Grassmann) symmetry.  The BRST approach has been used to quantize this model [12-14] but the local gauge symmetries associated with the BRST Hamiltonian has apparently not been considered.  The other model we shall look at is the first order Einstein-Hilbert action in $1 + 1$ dimensions.  The canonical structure of this model has been used to find a local gauge symmetry that is distinct from the manifest diffeomorphism invariance present in this model [15-17].  This novel symmetry has been used in conjunction with Faddeev-Popov path integral quantization [18]; we will consider the BRST approach to analyzing this model in section three.

\section{The Spinning Particle}
The action for a spinning particle is [11]
\begin{align}
S_C = \int d\tau \bigg[ \frac{1}{2} & \left(\frac{\dot{\phi}^2}{e} - i \psi \cdot \dot{\psi} - \frac{i}{e}\chi \dot{\phi} \cdot \psi \right) \nonumber \\
& + \frac{1}{2} \left( m^2 e + i\psi_5 \dot{\psi}_5 - i m \,\psi_5\, \chi \right)\bigg]
\end{align}
where $\phi^\mu$ and $e$ are Bosonic and $\psi^\mu$, $\psi_5$ and $\chi$ are Fermionic; $m$ is a mass parameter.  Both of the approaches of refs. [9, 10] lead to the generator of the gauge invariances associated with $S_C$ in eq. (7) being
\begin{align}
G = ( 2\dot{B} & + iF\chi )p_e + B\Theta \\
 &+ (-2i \dot{F})\pi_\chi + i F \Psi\,. \nonumber
\end{align}
In eq. (8), $p_A$ ($\pi_A$) is the canonical momentum associated with a Bosonic (Fermionic) coordinate variable $A$. Also,  
\begin{equation}
p_e  = \pi_\chi = 0
\end{equation}
are primary first class constraints,
\begin{equation}
\Theta \equiv p^2 - m^2 = 0
\end{equation}
\begin{equation}
\Psi \equiv p_\phi \cdot \psi - m\, \psi_5 = 0 
\end{equation}
are secondary first class constraints, $B(\tau)\, (F(\tau))$ is a Bosonic (Fermionic) gauge function, and the variables $\psi^\mu$, $\psi_5$ satisfy the DB
\begin{equation}
\left\lbrace \psi^\mu , \psi_\nu \right\rbrace^\ast = i\, \delta_\nu^\mu 
\end{equation}
\begin{equation}
\left\lbrace \psi_5 , \psi_5 \right\rbrace^\ast = -i 
\end{equation}
so that
\begin{equation}
\left\lbrace \Psi , \Psi \right\rbrace^\ast = i\,\Theta \,. 
\end{equation}
Following the procedure outlined in [1-7], it can be shown that the BRST operator $Q$ is given by 
\begin{equation}
Q = f_1 p_e + f_2 \Theta + b_1 \pi_\chi + b_2 \Psi + \frac{1}{2} \pi_{f_{2}} b_2^2 
\end{equation}
where $f_i(b_i)$ are Fermionic (Bosonic) ghost fields; subsequently it follows that the BRST Hamiltonian is 
\begin{equation}
H_Q = \frac{e}{2}\Theta  + \frac{i}{2} \chi \Psi - \frac{i}{2} f_1 \pi_{f_{2}} - \frac{1}{2} b_1 p_{b_{2}} + \frac{i}{2}\chi b_2 \pi_{f_{2}}
\end{equation}
\begin{equation}
\hspace{-5.2cm}= \left\lbrace Q, \Gamma \right\rbrace^\ast
\end{equation}
where
\begin{equation}
\Gamma = -\frac{i}{2} p_{b_{2}} \chi + \frac{i}{2} \pi_{f_{2}} e.
\end{equation}
(Eqs. (15-18) are also in refs. [12-14].)

With $H_Q$ given by eq. (16), we have the first order BRST action
\begin{align}
S_Q = & \int d\tau \bigg[ p_\phi \cdot \dot{\phi} + p_{b_{1}} \dot{b}_1 + 
p_{b_{2}} \dot{b}_2 - \frac{i}{2} \psi \cdot \dot{\psi} + \frac{i}{2} \psi_5\dot{\psi}_5 \\
& + i \pi_{f_{1}} \dot{f}_1 + i \pi_{f_{2}} \dot{f}_2 - H_Q \bigg].\nonumber
\end{align}
We now can perform a canonical analysis of the action $S_Q$.  It is apparent that again, there are the primary first class constraints of eq. (9).  Once more there is the secondary first class constraint of eq. (10), but now, in place of eq. (11), there is the secondary first class constraint
\begin{equation}
\overline{\Psi} \equiv \Psi + b_2 \pi_{f_{2}} = 0. 
\end{equation}
Since
\begin{equation}
\left\lbrace \overline{\Psi}, H_Q \right\rbrace^\ast = \frac{1}{2}\left( \chi \Theta - \pi_{f_{2}} b_1 \right) 
\end{equation}
there is now a tercery first class constraint
\begin{equation}
\pi_{f_{2}} b_1 = 0
\end{equation}
(recalling eq. (10)).  The formalism of ref. [10] can now be used to find the gauge generator associated with the local gauge invariances of $S_Q$ of eq. (19); it is 
\begin{equation}
G_Q = 2(\overline{B} + i \dot{\overline{F}} \chi ) p_e + \overline{B}\Theta - 4i \ddot{\overline{F}} \pi_\chi + 2i \dot{\overline{F}}\; \overline{\Psi} + i \overline{F} \pi_{f_{2}} b_1 ,
\end{equation}
where $\overline{B}(\overline{F})$ is a Bosonic (Fermionic) gauge function.  With $G_Q$, we find that $S_Q$ is invariant under the gauge transformations
\[\delta\chi = 4 \ddot{\overline{F}} \,,\; 
\delta \psi^\mu = 2\dot{\overline{F}} p_\phi^\mu \,, \;
\delta \psi_5 = 2m \dot{\overline{F}} \,, \;
\delta f_1 = 0 \,,   \;
\delta f_2 = -2\dot{\overline{F}} b_2 - \overline{F} b_1 
\]
\[ \delta p_e = \delta p_\phi^\mu = 0\,,\;
\delta p_{b_{2}} = -2i \dot{\overline{F}} \pi_{f_{2}}\,,\;
\delta p_{b_{1}} = -i \overline{F} \pi_{f_{2}} 
\]
\[\delta \pi_\chi = -2 \dot{\overline{F}} p_e\,,\;
\delta \pi_{f_{1}} = \delta \pi_{f_{2}} = 0.
\]
\[ \delta e = 2(\dot{\overline{B}} + i \dot{\overline{F}} \chi)\,,\;
\delta \phi^\mu = 2 \overline{B} p_\phi^\mu + 2i \dot{\overline{F}}
\psi^\mu 
\]
\[\delta b_1 = \delta b_2 = 0 ; \eqno(24a-p)\]  
under the transformations of eq. (24), it follows that
\[
\delta S_Q = \int d\tau \frac{d}{d\tau} \left[ \overline{B} (p_\phi^2 + m^2) + i \dot{\overline{F}} (2p_\phi \cdot \psi + m \psi_5)\right].\eqno(25)\]
Having established the presence of a local gauge symmetry in $S_Q$, we can now repeat the BRST procedure.  We first of all find that with eq. (2), the BRST operator associated with $S_Q$ is 
\[ \overline{Q} = \overline{f}_1 p_e + \overline{f}_2 \Theta +\overline{b}_1 \pi_\chi + \overline{b}_2 \overline{\Psi} + 
\overline{b}_3b_1 \pi_{f_{2}} + \frac{1}{2} \pi_{\overline{f}_{2}}\overline{b}_2^2 \eqno(26) \]
where $\overline{f}_i$ and $\overline{b}_i$ are Fermionic (Bosonic) ghost fields.  Again, from eq. (3), it follows that the BRST Hamiltonian that is associated with $\overline{Q}$ is 
\[ H_{\overline{Q}} = \frac{e}{2} \Theta + \frac{i}{2} \chi
\left( \Psi + b_2 \pi_{f_{2}} + \overline{b}_2 \pi_{\overline{f}_{2}}\right) - \frac{i}{2} f_1 \pi_{f_{2}}\nonumber\]
\[\hspace{2cm} - \frac{1}{2} b_1 p_{b_{2}} - \frac{i}{2} 
\overline{f}_1 \pi_{\overline{f}_{2}} - \frac{1}{2} \overline{b}_1 
p_{\overline{b}_{2}} + \frac{1}{2} \overline{b}_2 p_{\overline{b}_{3}}\,. \eqno(27) \]
We shall now examine the action
\[ S_{\overline{Q}} = \int d\tau \bigg [ p_\phi \cdot \dot{\phi} + p_{b_{1}}\dot{b}_1 + p_{b_{2}}\dot{b}_2  + p_{\overline{b}_{1}}\dot{\overline{b}}_1 +
p_{\overline{b}_{2}} \dot{\overline{b}}_2 + p_{\overline{b}_{3}} \dot{\overline{b}}_3 \nonumber \]
\[ \hspace{4cm}- \frac{i}{2} \psi \cdot \dot{\psi} + \frac{i}{2} \psi_5 \dot{\psi}_5 + i \pi_{f_{1}} \dot{f}_1  + i \pi_{f_{2}} \dot{f}_2
+ i \pi_{\overline{f}_{1}} \dot{\overline{f}}_1 + i \pi_{\overline{f}_{2}} \dot{\overline{f}}_2 - H_{\overline{Q}}\bigg]
 \eqno(28) \]
for possible gauge invariances.  Again employing the approach of ref. [10], we find that the generator of gauge symmetries that leaves $S_{\overline{Q}}$ in eq. (28) invariant is 
\[ \tilde{G} = 2(\tilde{B} - i \dot{\tilde{F}} \chi ) p_e + 
\tilde{B}\Theta + 4i\ddot{\tilde{F}} \pi_\chi - 2i\dot{\tilde{F}} (\Psi + b_2 \pi_{f_{2}} + \overline{b}_2 \pi_{\overline{f}_2}) + i \tilde{F} b_1 \pi_{f_{2}} - i \dot{\tilde{F}}\overline{b}_1 \pi_{\overline{f}_{2}},
\eqno(29) \]
with $\tilde{B}(\tilde{F})$ being a Bosonic (Fermionic) gauge function.  Since $S_{\overline{Q}}$ has a gauge symmetry, it too is subject to a BRST analysis.

\section{The First Order Einstein-Hilbert Action \\in $1 + 1$ Dimensions}

Another example of a gauge theory which has an associated BRST action that itself possesses a gauge symmetry is provided by the first order Einstein-Hilbert action in $1 + 1$ dimensions.  The classical action for this model is 
\[ S_C = \int d^2x \sqrt{-g} \,g^{\mu\nu} R_{\mu\nu}(\Gamma) \eqno(30) \]
where
\[ R_{\mu\nu} = \Gamma_{\mu\nu ,\lambda}^\lambda - 
\Gamma_{\lambda\mu , \nu}^\lambda + \Gamma_{\mu\nu}^\lambda \Gamma_{\sigma\lambda}^\sigma - \Gamma_{\sigma\mu}^\lambda \Gamma_{\lambda\nu}^\sigma \, . \eqno(31) \]
If now $h^{\mu\nu} = \sqrt{-g} \, g^{\mu\nu}$ and $G_{\mu\nu}^\lambda = \Gamma_{\mu\nu}^\lambda - \frac{1}{2}\left( \delta_\mu^\lambda \Gamma_{\sigma\nu}^\sigma + \delta _\nu^\lambda \Gamma_{\sigma\mu}^\sigma\right)$, then eq. (30) can be rewritten [15-18]

\[ S_C = \int d^2x \bigg[ \pi h_{,0} + \pi_1 h_{,0}^1 + \pi_{11} h_{,0}^{11} - \left( \xi^1 \phi_1 + \xi \phi + \xi_1 \phi^1 \right)\bigg]\eqno(32) \]
where
\[ h \equiv h^{00} \,, \quad h^1 \equiv h^{01} \eqno(33a,b) \]
\[ \pi = - G_{00}^0 \,, \quad \pi_1 = -2G_{01}^0 \,, \quad \pi_{11} = -G_{11}^0 \eqno(33c,d,e) \]
\[ \xi^1 =  G_{00}^1 \,, \quad \xi = 2G_{01}^1 \,, \quad \xi_{1} = G_{11}^1 \eqno(33f,g,h) \]
\[ \phi_1 = h_{,1} - h \pi_1 - 2 h^1\pi_{11} \eqno(33i) \]
\[ \phi = h_{,1}^1 + h \pi -  h^{11} \pi_{11} \eqno(33j) \]
\[ \phi^1 = h_{,1}^{11} + 2h^1 \pi +  h^{11} \pi_{1}\; . \eqno(33k) \]
The primary constraints
\[ p_{\xi^{1}} = p_\xi = p_{\xi_{1}} = 0 \eqno(34) \]
obviously lead to the secondary constraints
\[ \phi^1 = \phi = \phi_1 = 0 \; ; \eqno(35) \]
these are all first class as 
\[ \left\lbrace \phi , \phi^1 \right\rbrace = \phi^1 \; ,\quad
\left\lbrace \phi_1 , \phi \right\rbrace = \phi_1 \; ,\quad
\left\lbrace \phi_1 , \phi^1 \right\rbrace = 2\phi \; .\eqno(36a,b,c) \]
Using these constraints, one finds that the gauge generator leads to the transformations [15-18] 
\[ \delta h^{\alpha\beta} = - \left( \epsilon^{\alpha\mu} h^{\beta\nu} + \epsilon^{\beta\mu} h^{\alpha\nu} \right) \omega_{\mu\nu} \eqno(37a) \]
\[ \delta G_{\mu\nu}^\lambda = - \epsilon^{\lambda\rho} \omega_{\mu\nu , \rho} - \epsilon^{\rho\sigma} \left(G_{\mu\rho}^\lambda \omega_{\nu\sigma} + G_{\nu\rho}^\lambda \omega_{\mu\sigma}\right) \eqno(37b) \]
where $\epsilon^{01} = - \epsilon^{10} = 1$ and $\omega_{\mu\nu}$ is a symmetric gauge function.

If now we define
\[\phi^1 = \sigma_a + \sigma_c \;,\quad
\phi = \sigma_b \;,\quad
\phi_1 = -\sigma_a + \sigma_c \eqno(38)\]
\[ \xi_1 = \frac{1}{2} \left( \zeta_a + \zeta_c\right) \quad \xi = \zeta_b \quad \xi^1 = \frac{1}{2}\left( -\zeta_a + \zeta_c\right) \eqno(39) \]
then we have a canonical Hamiltonian
\[ H_C = \zeta_a\sigma_a + \zeta_b\sigma_b + \zeta_c\sigma_c \;.\eqno(40)\]
Introducing now Fermionic ghost fields $f_a$, $f_b$, $f_c$ and $F_a$, $F_b$, $F_c$, it follows from eqs. (2,3) that the BRST charge $Q$ and the BRST Hamiltonian $H_Q$ are given by
\[ Q = f_a p_{\zeta_{a}} + f_b p_{\zeta_{b}} + f_c p_{\zeta_{c}} + F_a\sigma_a + F_b\sigma_b + F_c\sigma_c \nonumber \]
\[ -i \pi_{F_{a}}F_b F_c  -i \pi_{F_{b}}F_c F_a +i \pi_{F_{c}}F_a F_b \eqno(41) \]
\[ H_Q = \zeta_a \Sigma_a + \zeta_b\Sigma_b + \zeta_c\Sigma_c + 
i \pi_{F_{a}}f_a + i \pi_{F_{b}}f_b + i \pi_{F_{c}}f_c   \eqno(42) \]
where
\[ \Sigma_a = \sigma_a - i\left(\pi_{F_{b}}F_c  + \pi_{F_{c}}F_b \right) \eqno(43a) \]
\[ \Sigma_b = \sigma_b + i\left(\pi_{F_{a}}F_c  + \pi_{F_{c}}F_a \right) \eqno(43b) \]
\[ \Sigma_c = \sigma_c - i\left(\pi_{F_{a}}F_b  + \pi_{F_{b}}F_a \right) \eqno(43c) \]
satisfy 
\[ \left\lbrace \Sigma_a ,\Sigma_b \right\rbrace = -\Sigma_c\;,\quad 
\left\lbrace \Sigma_b ,\Sigma_c \right\rbrace = \Sigma_a\;,\quad 
\left\lbrace \Sigma_c ,\Sigma_a \right\rbrace = \Sigma_b\;.
\eqno(44) \]
We find that
\[ H_Q = \left\lbrace Q ,\Gamma\right\rbrace \eqno(45) \]
where
\[ \Gamma = i\left( \zeta_a F_a + \zeta_b F_b + \zeta_cF_c\right).\eqno(46) \]
The action associated with the BRST Hamiltonian $H_Q$
\[ \hspace{-2cm}S_Q = \int d^2x \bigg[ \pi h_{,0} + \pi_1 h_{,0}^1 + \pi_{11} h^{11}_{,0} + i \big( \pi_{f_{a}} f_{a,0} + \pi_{f_{b}} f_{b,0} \nonumber \]
\[+ \pi_{f_{c}} f_{c,0} + \pi_{F_{a}} F_{a,0} + \pi_{F_{b}} F_{b,0} + \pi_{F_{c}} F_{c,0} \big) - H_Q \bigg] \eqno(47) \]
obviously has the primary constraints
\[ p_{\zeta_{a}} = p_{\zeta_{b}} = p_{\zeta_{c}} = 0 \eqno(48) \]
as well as the secondary constraints
\[ \Sigma_a = \Sigma_b = \Sigma_c = 0 \, ; \eqno(49) \]
with $H_Q$ given by eq. (42) there are no tertiary constraints. The constraints of eqs. (48, 49) are all first class and consequently $S_Q$ itself possesses a gauge invariance which is generated by [10]
\[ G_Q = \left( \dot{B}_a + B_b \zeta_c - B_c\zeta_b \right)p_{\zeta_{a}} + \left( \dot{B}_b + B_c \zeta_a - B_a\zeta_c \right)p_{\zeta_{b}} \eqno(50) \]
\[ \hspace{3cm}+ \left( \dot{B}_c - B_a \zeta_b + B_b\zeta_a \right)p_{\zeta_{c}} + 
B_a \Sigma_a + B_b \Sigma_b + B_c \Sigma_c \nonumber \]
where $B_a$, $B_b$, $B_c$ are gauge functions.

\section{Discussion}
Cancellation of the effects due to the presence of non-physical degrees of freedom appearing in a locally gauge invariant actions by the introduction of ``ghost'' fields is quite efficient.  It is well understood that the BRST action of eq. (4) that takes the place of the classical action of eq. (5) upon introduction of these ghost fields has a global gauge invariance on account of eq. (3) [6]; we have in this paper demonstrated that the BRST action itself might possess a local gauge invariance.  Adding a term of the form 
\[ H_{gf} = \left\lbrace Q, \Gamma^\prime \right\rbrace^\ast \eqno(51) \]
to $H_Q$ is a form of ``gauge fixing'' [1-7].  This term leaves eq. (3) intact on account of eq. (2) and it has been demonstrated [1-7] that transition amplitudes are independent of the choice of $\Gamma^\prime$; it may also break any local symmetry present in $S_Q$.  We note that on account of eqs. (17, 45) $H_Q + H_{gf} = 0$ for both the spinning particle and the Einstein-Hilbert action in $1 + 1$ dimension if we choose $\Gamma^\prime = -\Gamma$.  In the discussion of the spinning string in refs. [1-14], $\Gamma^\prime = 0$.  In this case, since the BRST action has a local gauge invariance, the path integral used in quantization is not well defined.  One could either choose a suitable gauge fixing function $\Gamma^\prime$ or reapply the BRST procedure and after having arrived at an action involving ``ghosts of ghosts'', check to see if it is well defined; if it is not, a gauge fixing function can be introduced at this stage.

The novel ``ghosts of ghosts'' arising due to the situation described above differ from the ``new ghosts'' that may arise in the course of applying the BV analysis of ref. [19].  These new ghosts of BV arise whenever the gauge invariance present in the Lagrangian has reducible generators; they serve to eliminate any gauge invariance that would be present in the ghost sector if they were not included.  In contrast, the ``ghosts of ghosts'' that are considered in this paper may occur even if the gauge generators of the classical action are not reducible; their purpose is to ensure that if the BRST action that arises upon applying the procedure of refs. [1-3] itself possesses a gauge invariance then any superfluous degrees of freedom occurring are eliminated in a consistent way.  We note that we had not expected that the BRST action might itself be gauge invariant; in Yang-Mills theory this is not the case.  It would appear that each BRST action must be examined individually to see if it is gauge invariant using the standard approach of refs. [9,10].

We note that the BRST approach of refs. [1-3] and the BV approach of ref. [19] are related.  A discussion of their connection appear in ref. [20] where the gauge invariance present in the first order (Hamiltonian) form of the action is treated using both approaches and they are shown to be equivalent.

\section*{Acknowledgements}
Roger Macleod has a helpful suggestion.

\section*{Appendix A Conventions}
For Grassmann variables $\theta_A$, we have
\[ \left(\theta_A \theta_B \right)^\dagger = \theta_B^\dagger \theta_A^\dagger \eqno(A.1) \]
and use left derivatives so that
\[ \frac{d}{d\theta_A }\left(\theta_B \theta_C \right) = \delta_{AB} \theta_C - \delta_{AC} \theta_B \; .\eqno(A.2) \]
If $(q_i, p_i)$ and $(\theta_i , \pi_i)$ are canonically conjugate pairs of standard and Grassmann variables respectively than for a theory with Lagrangian $L(q_i, \theta_i; \;\dot{q}_i, \dot{\theta}_i)$ we have
\[ p_i = \frac{\partial L}{\partial \dot{q}_i} \qquad 
\pi_i = i \frac{\partial L}{\partial \dot{\theta}_i} \eqno(A.3,4) \] 
and define the canonical Hamiltonian
\[ H_c = p_i \dot{q}_i + i \pi_i \dot{\theta}_i - L \;.\eqno(A.5) \]
For standard and Grassmann quantities $B_i$ and $F_i$ respectively we use the Poisson Brackets (PB) 
\[ \left\lbrace B_1, B_2 \right\rbrace = \left( B_{1,q} B_{2,p} - B_{1,p} B_{2,q} \right) + i \left(B_{1,\theta} B_{2,\pi} + B_{1,\pi} B_{2,\theta} \right) \eqno(A.6a)\]
\[ \left\lbrace F_1, F_2 \right\rbrace = \left( F_{1,q} F_{2,p} - F_{1,p} F_{2,q} \right) - i \left(F_{1,\theta} F_{2,\pi}+ F_{1,\pi} F_{2,\theta} \right) \eqno(A.6b)\]
\[ \left\lbrace F, B\right\rbrace = \left( F_{,q} B_{,p} - F_{,p} B_{,q} \right) - i \left(F_{,\theta} B_{,\pi} + F_{,\pi} B_{,\theta} \right)= \left\lbrace B, F \right\rbrace \;. \eqno(A.6c)\]
These conventions are consistent with those of ref. [6].

\end{document}